\documentclass[aps, prd, preprint, superscriptaddress, showpacs]{revtex4}
\usepackage{graphicx}
\usepackage{ulem}
\usepackage{epsfig}
\usepackage{color}
\usepackage{multirow}
\usepackage{epstopdf}
\definecolor{My_red}        {cmyk}{0.00, 1.00, 1.00, 0.20}

\newcommand{\bmat}{\left(\begin{array}}
\newcommand{\emat}{\end{array}\right)}
\newcommand{\beq}{\begin{equation}}
\newcommand{\eeq}{\end{equation}}

\newcommand{\lsim}{\mathrel{\ltap}}

\usepackage[centertags]{amsmath}
\usepackage{amssymb}
\usepackage{hyperref}

\newcommand{\missingET}{\mathchoice{\rlap{\kern.2em/}E_T}{\rlap{\kern.2em/}E_T}{\rlap{\kern.1em$\scriptstyle/$}E_T}{\rlap{\kern.1em$\scriptscriptstyle/$}E_T}}

\let\jnfont=\rm
\def\NPB#1,{{\jnfont Nucl.\ Phys.\ B }{\bf #1},}
\def\PLB#1,{{\jnfont Phys.\ Lett.\ B }{\bf #1},}
\def\EPJC#1,{{\jnfont Eur.\ Phys.\ Jour.\ C }{\bf #1},}
\def\PRD#1,{{\jnfont Phys.\ Rev.\ D }{\bf #1},}
\def\PRL#1,{{\jnfont Phys.\ Rev.\ Lett.\ }{\bf #1},}
\def\MPLA#1,{{\jnfont Mod.\ Phys.\ Lett.\ A }{\bf #1},}
\def\JPG#1,{{\jnfont J.\ Phys.\ G }{\bf #1},}
\def\CTP#1,{{\jnfont Commun.\ Theor.\ Phys.\ }{\bf #1},}
\def\JHEP#1,{{\jnfont JHEP \ }{\bf #1},}
\def\NPPS#1,{{\jnfont Nucl.\ Phys.\ Proc.\ Suppl.\ }{\bf #1},}
\def\CPC#1,{{\jnfont Computl.\ Phys.\ Commun.\ }{\bf #1},}
\def\CPL#1,{{\jnfont Chin.\ Phys.\ Lett. }{\bf #1},}
\def\APPB#1,{{\jnfont Acta\ Phys.\ Polon.\ B }{\bf #1},}

\def\lsim{\raise0.3ex\hbox{$<$\kern-0.75em\raise-1.1ex\hbox{$\sim$}}}
\def\gsim{\raise0.3ex\hbox{$>$\kern-0.75em\raise-1.1ex\hbox{$\sim$}}}

\begin{document}
\preprint{TTP14-031}

\title{Pseudo-goldstino and electroweakinos via VBF processes at LHC}

\author{Tao Liu}
\affiliation{Institut f\"ur Theoretische Teilchenphysik, Karlsruhe
  Institute of Technology (KIT), D-76128 Karlsruhe, Germany}

\author{Lin Wang}
\affiliation{Institut f\"ur Theoretische Teilchenphysik, Karlsruhe
  Institute of Technology (KIT), D-76128 Karlsruhe, Germany}

\author{Jin Min Yang}
\affiliation{State Key Laboratory of Theoretical Physics,
      Institute of Theoretical Physics, Academia Sinica, Beijing 100190,
      China \vspace{2cm}}

\begin{abstract}
The multi-sector SUSY breaking predicts pseudo-goldstino
which can couple to the visible sector more strongly than the ordinary
gravitino and thus induce the decays of the lightest neutralino and chargino
(collectively called electroweakinos) inside the detector.
In this note we study the electroweakino pair productions
via VBF processes followed by decays to pseudo-goldstino at the LHC.
Our Monte Carlo simulations show that
at the 14 TeV LHC with 3000 $fb^{-1}$ luminosity
the dominant production channel $pp\to \chi_1^{\pm} \chi_1^{0} jj$
can have a statistical significance above $2\sigma$ while other production
channels are not accessible.

\end{abstract}

\maketitle

\section{INTRODUCTION}
Search for supersymmetry (SUSY) is an important task for the
LHC. The current null search results indicate that the SUSY breaking scale
may be far above the weak scale and hence we must tolerate some extent of
unnaturalness. However, for the explanation of dark matter relic density
and the unification of gauge couplings, the electroweak gauginos
and higgsinos (collectively called electroweakinos) cannot be too heavy
and should be accessible in the upcoming runs of the LHC \cite{wenyu}.
Search for these electroweakinos at the LHC is rather challenging and has been
recently intensively studied \cite{gauginos-LHC}.
For example, when the lightest electroweakinos have compressed
mass spectrum, their pair productions through Drell-Yan processes
only give missing energy and an extra jet or gauge boson is needed
for the detection \cite{monojet}.
Another type of productions of  electroweakinos at the LHC
is vector boson fusion (VBF), which is shown quite promising
despite of small cross sections \cite{VBF-gauginos}.
These VBF productions naturally produce two highly energetic quark jets
with large dijet invariant mass in the forward and backward regions
of the detector \cite{Bjorken:1992er}.
An important feature of VBF is the absence of color exchange between
these two jets, which leads to a reduction of gluon emission
in the central region. This is in contrast to the case of typical
QCD backgrounds. Due to this feature, the VBF processes have been studied for
producing  electroweakinos \cite{VBF-gauginos}
and the Higgs bosons \cite{VBF-HIGGS}.

Note that in the VBF productions of electroweakinos at the LHC, e.g., the dominant channel
$pp\to \chi^\pm_1 \chi^0_1 jj$, in order to have a sizable cross
section, the lightest electroweakinos ($\chi^\pm_1$ and $\chi^0_1$) must be
wino-like and have compressed mass spectrum, which gives a signal of
two jets plus missing energy in the general framework of minimal supersymmetric model
with R-parity.
In the multi-sector SUSY breaking scenario, however,
both $\chi^\pm_1$ and $\chi^0_1$ can decay to visible particles
plus pseudo-goldstino inside the detector and then the VBF production
$pp\to \chi^\pm_1 \chi^0_1 jj$ will give rather different signals.
Such multi-sector SUSY breaking scenario refers to SUSY breaking in more than
one hidden sector, in which one goldstino will become the longitudinal component
of gravitino and other orthogonal states will become the physical
pseudo-goldstinos. Unconstrained by the supercurrent,
the couplings of the pseudo-goldstinos could be large enough to have
intriguing phenomenology
\cite{Cheung:2010mc,Cheung:2010qf,Benakli:2007zza,Craig:2010yf,
McCullough:2010wf,Izawa:2011hi,Thaler:2011me,Cheung:2011jq,Bertolini:2011tw,
Cheng:2010mw,Argurio:2011hs,Mawatari:2011jy,Argurio:2011gu,Ferretti:2013wya,
Liu:2013sx}. In our previous work
\cite{Hikasa:2014yra} we investigated the Drell-Yan productions of
the lightest electroweakinos followed by the
decays to pseudo-goldstino at the LHC.
In this work we extend the study to the VBF productions
of  electroweakinos.

The structure of this note is as follows. In Section II, we briefly
describe the neutralino and chargino decays to pseudo-goldstino, and
then perform the Monte Carlo simulations for the signal and backgrounds
of their VBF productions at the LHC.
Finally our conclusions are given in Section III.

\section{Phenomenological study at the LHC}
\subsection{Chargino/neutralino decays to pseudo-goldstino}
\begin{figure}[htbp]
\includegraphics[scale=0.8]{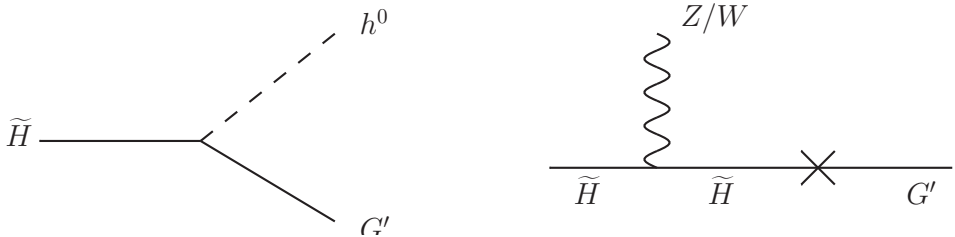}
\caption{A diagrammatic description of the interaction
and mixing between pseudo-goldstino $G^\prime$
and higgsino $\tilde{H}$ in the two-sector SUSY breaking scenario.}
\label{dia1}
\end{figure}

We now recapitulate our scenario (for a detailed description, see our previous work \cite{Hikasa:2014yra}).
Our scenario is that SUSY is broken in two hidden sectors, in which
the sector with a low SUSY breaking scale gives very small contribution
to the soft gaugino masses.  In this scenario the pseudo-goldstino ($G^{\prime}$) couplings
to the photon and transverse $Z$-boson are suppressed while its interaction with
Higgs and longitudinal gauge bosons are enhanced comparing to ordinary gravitino.
So a light neutral higgsino can decay to a Higgs boson or $Z$-boson plus $G^{\prime}$
while a light charged higgsino can only decay to a $W$-boson plus $G^{\prime}$ (because
the charged Higgs is usually heavier than the light charged higgsino), as shown
in Fig.~\ref{dia1}.
In the VBF productions, the chargino ($\chi^+_1$) and  neutralino  ($\chi^0_1$)
must be wino-like in order for a sizable production rate. Such a wino-like  neutralino
can have a decay $\chi^0_1\to h + G'$ through its mixing with the neutral
higgsino, while the wino-like chargino can have a decay  $\chi^+_1\to W + G'$
through two insertions.
Since the two insertions may lead to a rather small decay width,
the gravitino (goldstino) channel $\chi^+_1\to W + G$ can be
comparable. Considering the gravitino and  pseudo-goldstino have
the same collider signature (missing energy),
we only consider the pseudo-goldstino decay channel.

As in our previous work \cite{Hikasa:2014yra}, we can start the analysis
from the effective interaction between pseudo-goldstino,
chargino and neutralino
\begin{align}
\mathcal{L}_{\rm eff}=\dfrac{\widetilde
m_\phi^2}{F}[g_{h\chi}h\chi^{0} G^{\prime}+g_{\chi
Z}\bar{G^{\prime}}\bar{\sigma}^{\mu}\chi^{0} Z_{\mu} + g_{\chi
W_{1}} \bar{G^{\prime}} \bar{\sigma}^{\mu} \chi^{+} W^{-}_{\mu} +
g_{\chi W_{2}} \bar{G^{\prime}} \bar{\sigma}^{\mu} \chi^{-}
W^{+}_{\mu} +h.c.], \label{eff}
\end{align}
where $F=\sqrt{F_1^2+F_2^2}$ with $F_i$ being the SUSY breaking scales in
two hidden sectors, $\widetilde{m}^2_{\phi}=-m^2_{\phi,1} \tan\theta+m^2_{\phi,2} \cot\theta$
with $\tan\theta=F_2/F_1$ and $m_{\phi,i}$ the soft masses for the chiral fields.
With fixed parameters $\widetilde m_\phi/\sqrt{F}=0.1$ and all the couplings $g_X=1$,
the decay width of neutralino or chargino into pseudo-goldstino is of the order
$\sim 10^{-4}$ GeV and the decay length is about $10^{-10}$ cm which means that the
decays will occur inside the detector.
As shown in Fig.~2 of \cite{Hikasa:2014yra}, the lightest electroweakinos
decay dominantly to pseudo-goldstino and in our calculation we the assume
such a decay has a branching of 100\%.

\subsection{Signal of VBF productions of chargino/neutralino }
\begin{figure}[htbp]
\includegraphics[scale=0.7]{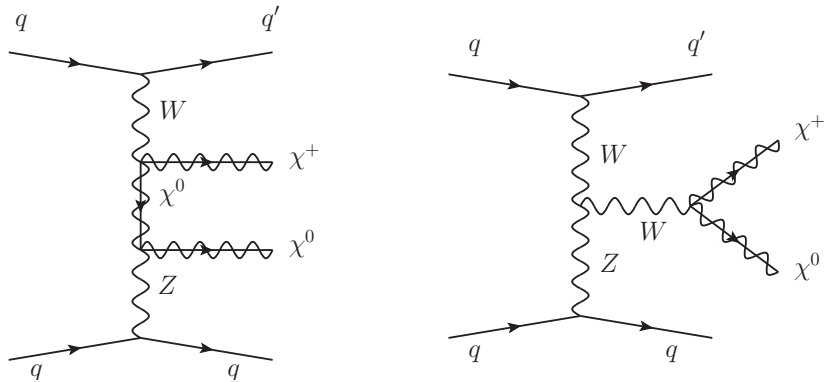}
\caption{The representative Feynman diagrams for
neutralino and chargino pair production via VBF processes
at the LHC.} \label{dia2}
\end{figure}

For neutralino/chargino
productions through VBF processes, we focus on the pair production of
a neutralino and a chargino. The representative Feynman diagrams are shown
in Fig.~\ref{dia2}. Note that apart from these pure VBF processes, some non-VBF processes
could also provide the same final states. For instance, the higher order QCD
effects in the Drell-Yan productions of neutralino and chargino
could also give contributions because of the hard emission of partons
from the initial states. In our calculation, we consider the full set of diagrams
and employ kinematic constraints to reduce the contribution from non-VBF processes.

The main feature of VBF processes is the presence of two jets
produced in the forward and backward regions at the detector and with
large pseudo-rapidity separation between them. They also must be hard
enough in order to create a pair of neutralino and chargino.
Therefore, we calculate
the cross sections by characterizing the signal in terms of the
following selections:
\begin{itemize}
 \item[(a)] The two jets in the forward/backward regions, labeled as $j_{1}$ and $j_{2}$,
 must satisfy the requirement
 $\mid \Delta \eta (j_1,j_2) \mid > 4.2$ and $\eta_{j_1}\cdot\eta_{j_2} < 0$.
 \item[(b)] We accept the jets with $P_{T}^{j_1,j_2} > 40$ GeV and $\mid \eta_j \mid < 5$.
 \item[(c)] The invariant mass of these two jets should be large, $M_{j_1j_2}>500$ GeV.
\end{itemize}

\begin{figure}[htbp]
\includegraphics[scale=0.8]{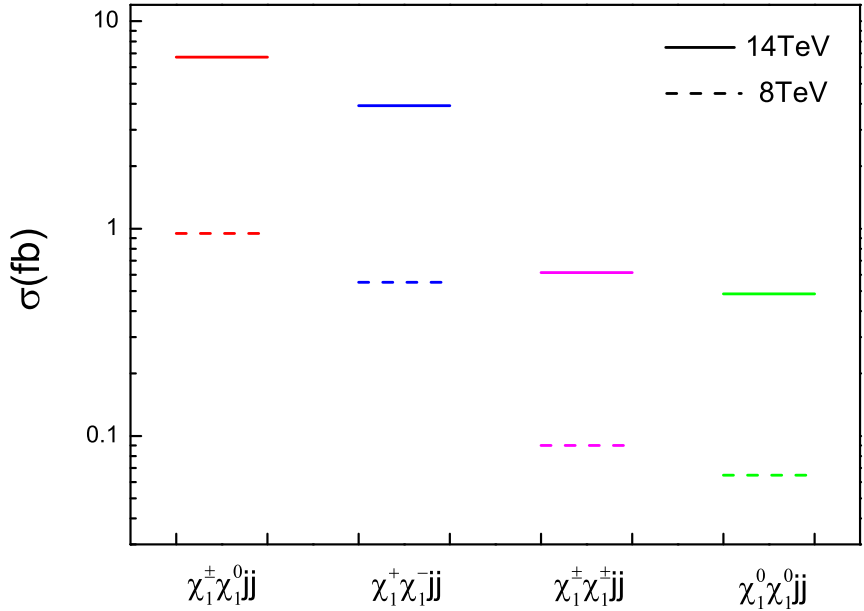}
\vspace*{-0.9cm}
\caption{The cross section for neutralino and
chargino production in association with two jets after imposing
selections (a)-(c) at the LHC with $\sqrt{s}=$ 14 TeV and 8 TeV.}
\label{dia3}
\end{figure}

Now we discuss the productions of neutralinos and charginos through VBF.
In our study,  the squarks, sleptons, gluino and non-SM Higgs bosons are assumed
too heavy to be inaccessible. We fix the mass of the SM-like Higgs boson at 125 GeV.
For the parameter space in the neutralino/chargino sector, we focus on the region
in which the lightest neutralino and chargino are wino-like because only the
wino-like neutralino and chargino can be sizably produced via the VBF processes
at the LHC \cite{VBF-gauginos}.

For the wino-like neutralino and chargino, they are produced through the VBF channel
\begin{equation}
 p p \rightarrow \chi_1^\pm \chi_1^0 j j,~~\chi_1^+ \chi_1^- j j,~~
 \chi_1^\pm \chi_1^\pm j j,~~\chi_1^0 \chi_1^0 j j.
\end{equation}
In our numerical calculation we choose the same benchmark scenario as in \cite{Hikasa:2014yra}:
\begin{equation}
M_{2} = 200 ~{\rm GeV}, ~~\mu = 1.0 ~{\rm TeV},  ~~M_1 = 1.5  ~{\rm TeV}, ~~\tan\beta = 10.
\end{equation}
Note that our results are not sensitive to the value of $\tan\beta$
because the wino-like neutralino and chargino are produced in VBF processes
dominantly via the gauge couplings. The value of $\tan\beta$ can  
only have effects through the higgsino component which is small 
in a wino-like neutralino or chargino. 
On the other hand, 
since the neutralino and chargino produced in VBF processes are wino-like,
their masses are mainly determined by the value of $M_2$. So for a lower value 
of $M_2$ (while being consistent with current LHC and LEP limits),
the neutralino and chargino are lighter, 
whose production rate is larger and the statistical significance 
can be higher.

We first calculate their cross sections at the partonic level using
the package MadGraph5 \cite{MG5} and employ these cuts (a)-(c) for the two jets.
The results at the LHC with $\sqrt{s} = 14$ TeV and
$ 8 $ TeV are displayed in Fig.~\ref{dia3}. We
check our results with CalcHEP \cite{CalcHEP} and find agreements.
The results in Fig.~\ref{dia3} show that the largest cross sections come
from $pp\to \chi_1^{\pm} \chi_1^0 j j$ and $pp\to \chi_1^{+} \chi_1^{-} j j$.
The cross sections at 8 TeV LHC are approximately 7-8 times smaller than 14 TeV LHC.
Therefore, we consider the these two channels
at the 14 TeV LHC in the following analysis.  We also check
the higgsino-like and bino-like neutralino/chargino and find that their production
cross sections are much smaller than the wino-like case.

\subsection{The observability of $\chi_1^\pm \chi_1^0 j j$ production at the LHC}
First we focus on the $\chi_1^\pm \chi_1^0 j j$ production.
As we discussed earlier, in our scenario the lightest chargino
$\chi^{\pm}_1$ decays to a pseudo-goldstino plus a $W$-boson
and  the lightest neutralino  $\chi^0_1$ decays to a pseudo-goldstino
plus a Higgs boson. Thus the signal of this production
is a single lepton and two bottom quarks associated with
two energetic light jets and large missing transverse energy:
\begin{equation}
 p p \rightarrow \chi_1^\pm \chi_1^0 j j \rightarrow
 W^\pm h G^\prime G^\prime j j \rightarrow
 \ell^\pm \nu b \bar{b} G^\prime G^\prime j j \rightarrow
 \ell + 2b  + 2j + \missingET  ~~(\ell = e,\mu, \tau).
\end{equation}
The dominant SM backgrounds are from the production of top quark pair
with semi-leptonic decays of top quarks. The di-leptonic decays of
top pair could also fake the signal when the $\tau$ lepton
decays hadronically. In addition, the single top production
and $W$+jets may also mimic the signal.
The jets from these backgrounds are less energetic and
more central in the detector, which are different from the signal.
Therefore, the VBF selection cuts could reduce them effectively.
The productions of $WV$ ($V=W,Z$) via
VBF processes with the $W$-boson decaying leptonically and
the vector boson decaying to a pair of quarks could also fake the signal.
The missing energy in all these backgrounds come from
neutrinos. But for the signal process, the pseudo-goldstino
$G^{\prime}$ escapes the detector and leads to large missing energy.
So the $\missingET$ cut could further reduce these backgrounds.
Besides, the top pair production associated with a $Z$-boson and
the production of $Wh$ via VBF process could also fake the signal.
Due to smaller cross sections than other backgrounds, we do not
consider them in this work.

We use MadGraph5 \cite{MG5} to generate the signal and background events.
For our signal events generation, the effective Lagrangian in
Eq.~\ref{eff} is implemented in FeynRules \cite{Feynrules} and then passed to MadGraph5
via the UFO model file \cite{Ufo}. We apply Pythia \cite{Pythia} for
parton shower and hadronization, Delphes \cite{Delphes} with the ATLAS detector
for the fast detector simulations. The MLM scheme \cite{MLM} is used to
match our matrix element with parton shower. Jets are clustered employing
FastJet \cite{Fastjet} with anti-$k_{t}$ algorithm \cite{anti-kt} using the
radius parameter $\Delta R = 0.5$. Finally, we employ MadAnalysis5
\cite{madanalysis5} to perform sample analysis.

\begin{figure}[htbp]
\includegraphics[scale=0.4]{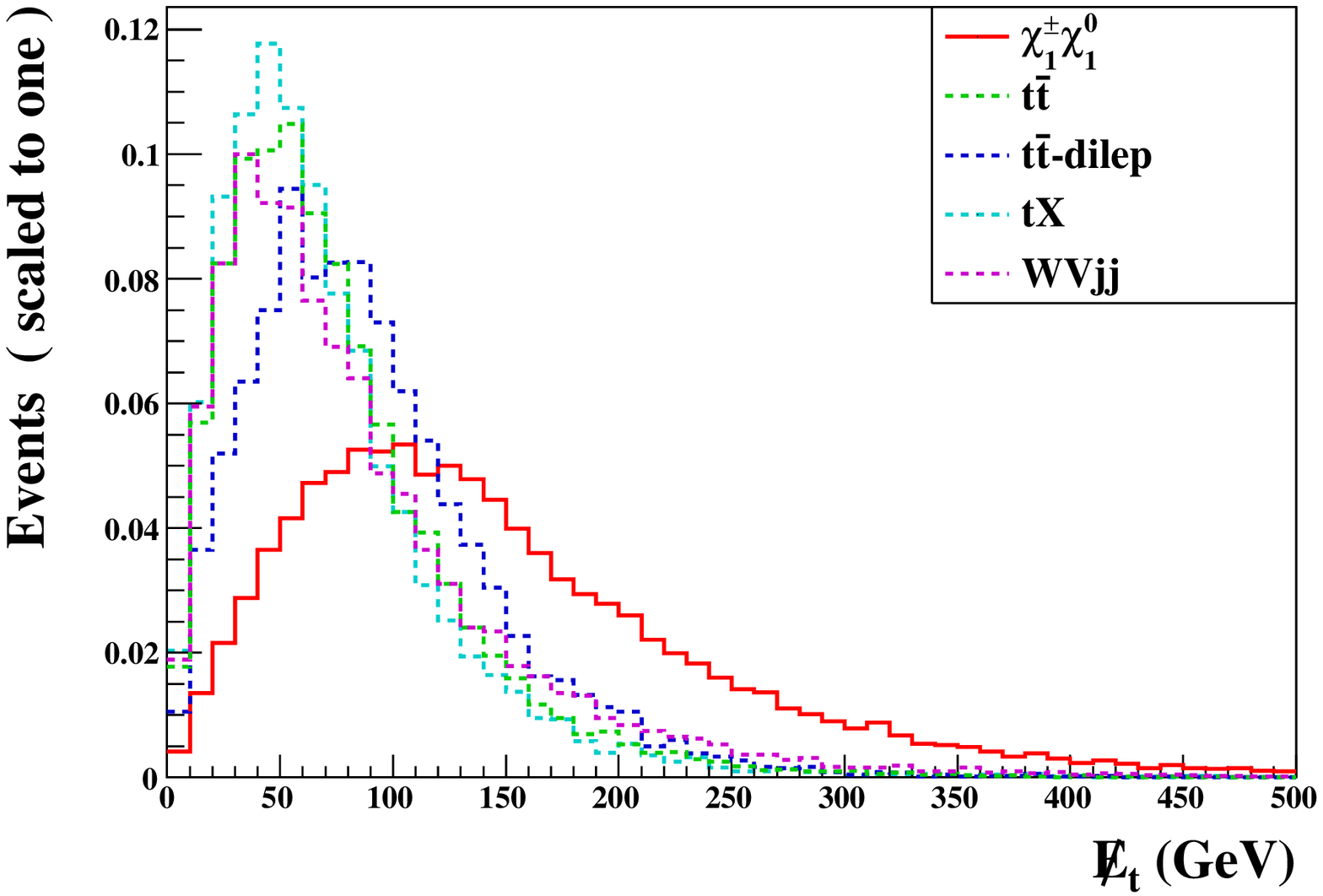}%
\includegraphics[scale=0.4]{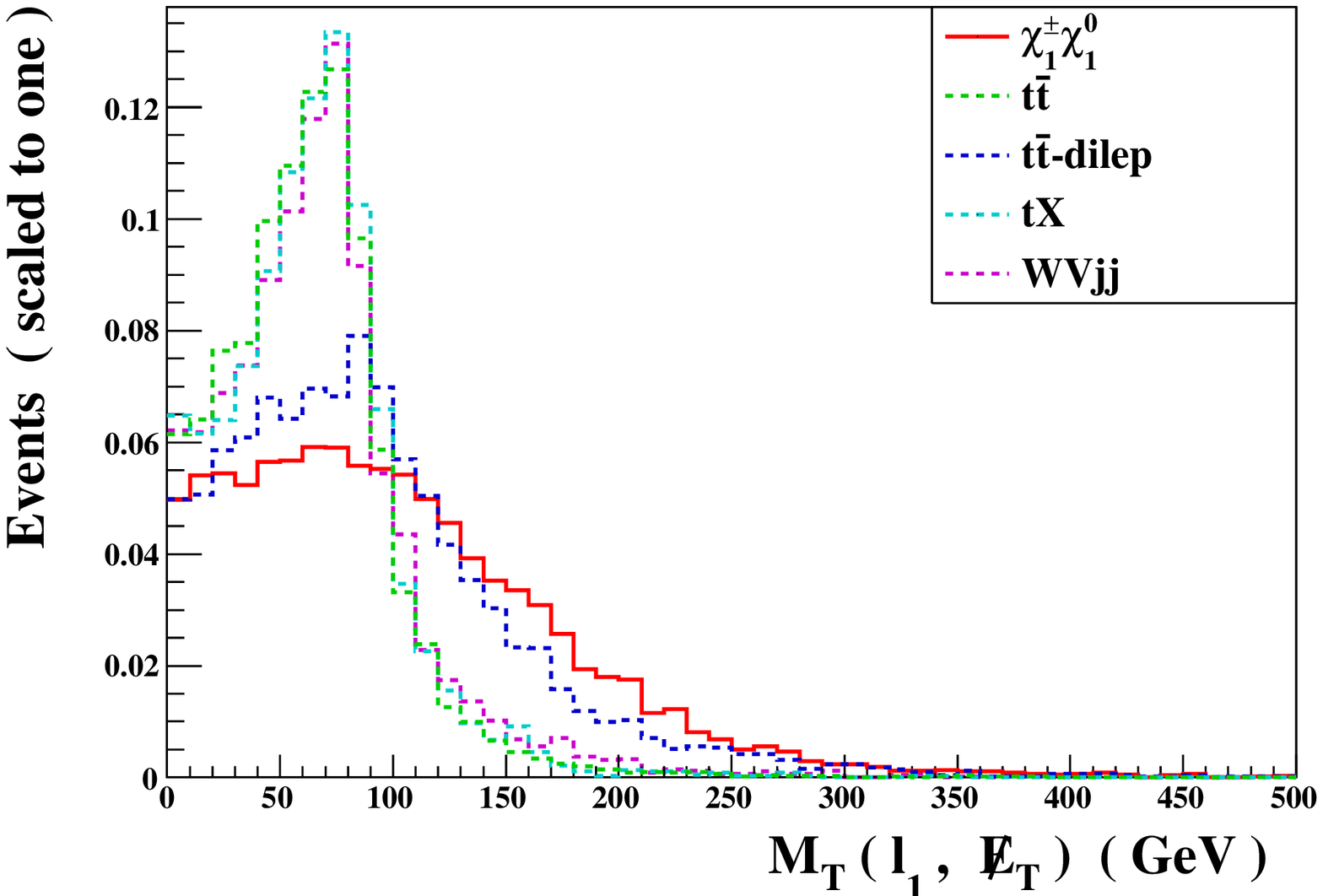}%
\vspace*{-0.3cm}
\caption{The normalized $M_{T}$ and $\missingET$ distributions
for the signal  $p p \rightarrow \chi_1^\pm \chi_1^0 j j \rightarrow
 W^\pm h G^\prime G^\prime j j \rightarrow
 l^\pm \nu b \bar{b} G^\prime G^\prime j j \rightarrow
 l + 4j + \missingET $
and background processes after VBF selections at the LHC
with $\sqrt{s}=14$~TeV.
For the signal we fixed the relevant mass parameters
as $\mu=200$~GeV, $M_{1}=1.0$~TeV, $M_{2}=1.5$~TeV.}
\label{figetmt}
\end{figure}

\begin{figure}[htbp]
\includegraphics[scale=0.4]{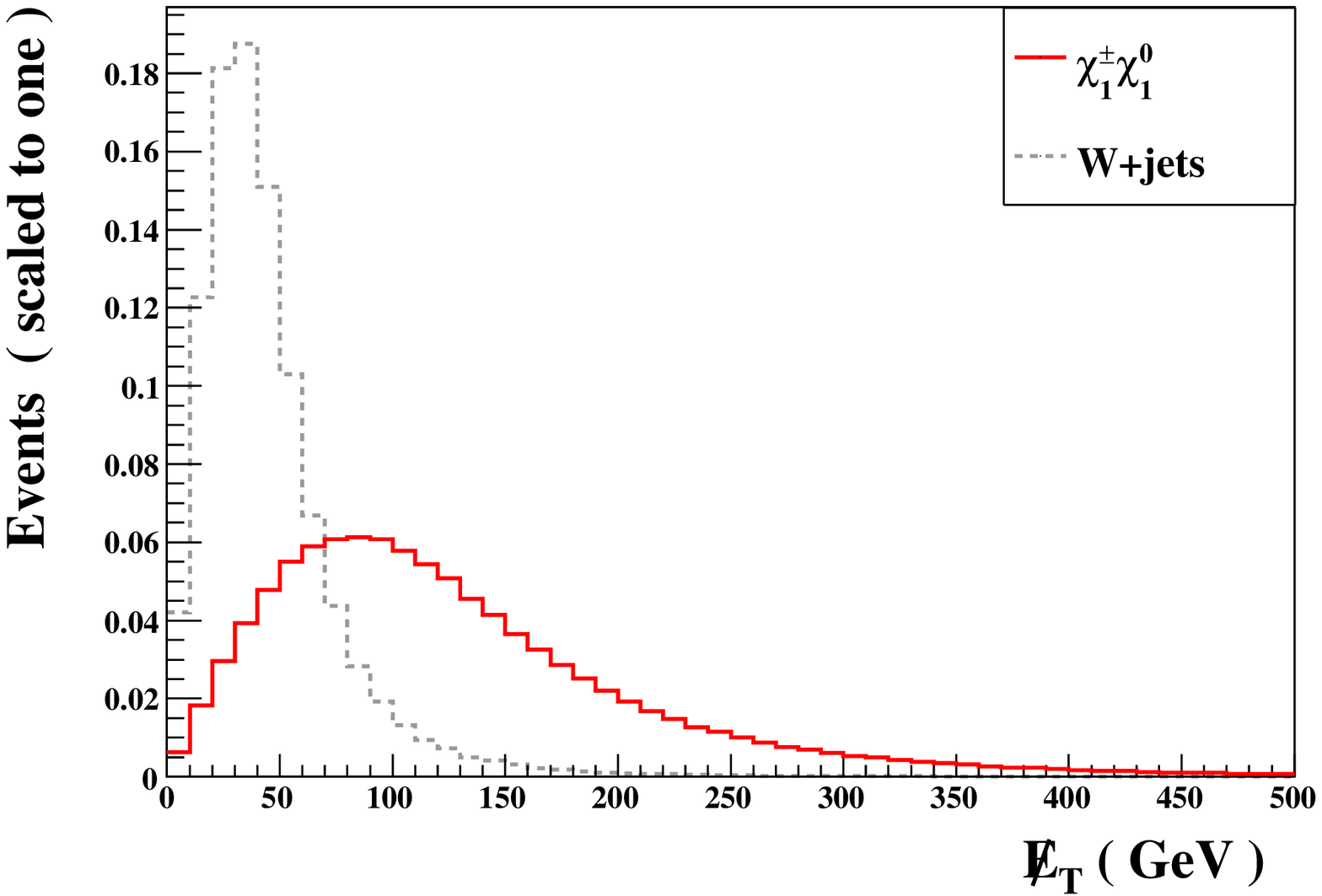}%
\includegraphics[scale=0.4]{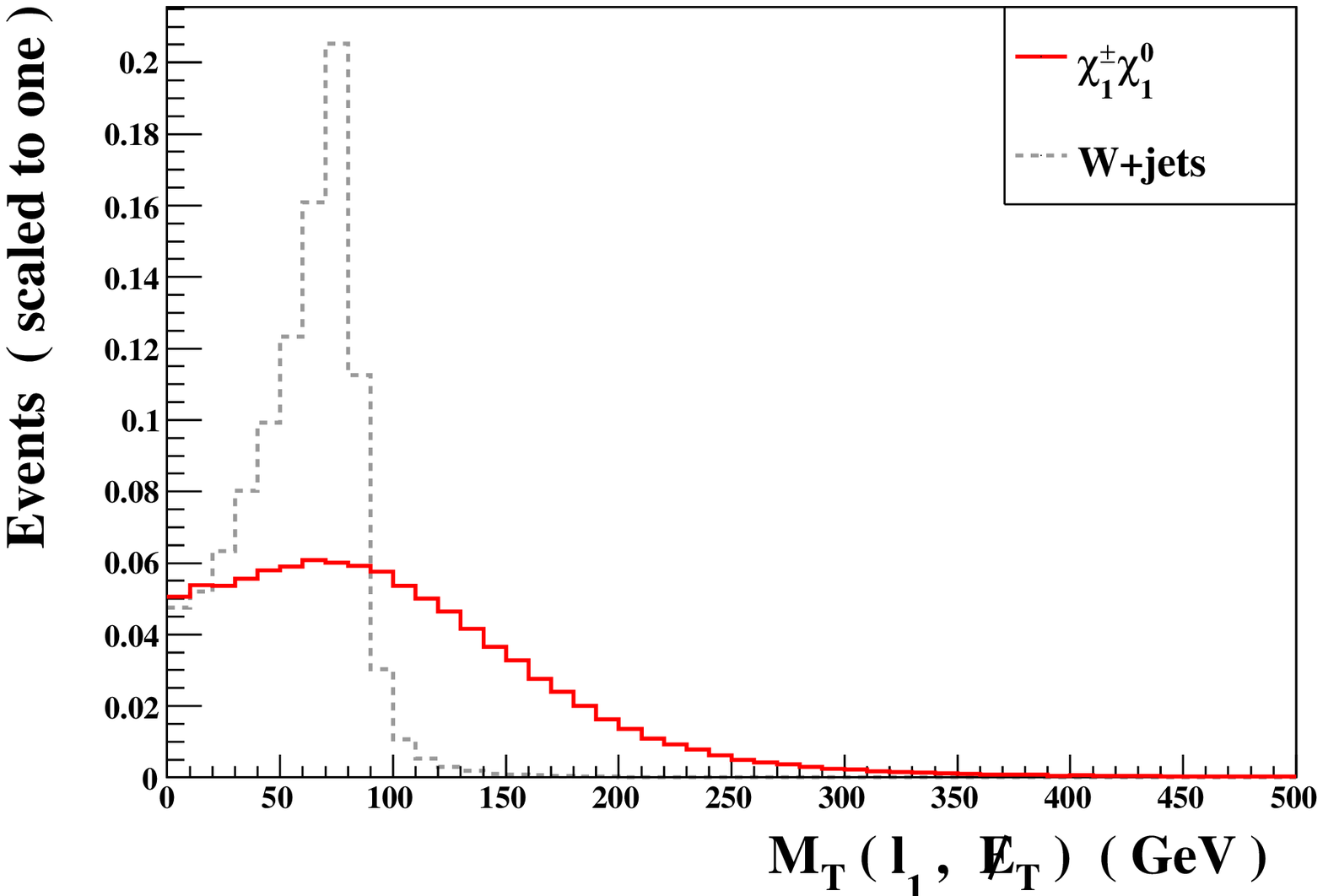}%
\vspace*{-0.3cm}
\caption{The normalized $M_{T}$ and $\missingET$ distributions
for the signal $p p \rightarrow \chi_1^\pm \chi_1^0 j j \rightarrow
 W^\pm h G^\prime G^\prime j j \rightarrow
 l^\pm \nu b \bar{b} G^\prime G^\prime j j \rightarrow
 l + 4j + \missingET $
and $W$+jets before VBF selections at the LHC with $\sqrt{s}=14$~TeV.
}
\label{figwjj}
\end{figure}

As we discussed earlier, we must first impose
the VBF selections to make the VBF processes dominant. Then
we present some kinematic distributions in order to get some
other efficient cuts.
In Fig.~\ref{figetmt}, we display the normalized distribution
of $\missingET$ and transverse mass $M_{T}(l_1,\missingET)$
for the signal and background processes at the LHC with $\sqrt{s}=14$ TeV,
where the $M_{T}(l,\missingET)$ is defined as
\begin{equation}
M_{T}=\sqrt{2p_{T}^{\ell} \missingET [1 - \cos\Delta\phi_{\ell,\missingET}]},
\end{equation}
with $\Delta\phi_{\ell,\missingET}$ standing for the azimuthal angle
difference between the lepton and the missing energy.
These distributions show that requiring a lower cut of about
150 GeV for $\missingET$ and 100 GeV for $M_{T}(l_1,\missingET)$ could be
effective to reduce the backgrounds. Note that we have checked
the VBF selection efficiency is $10^{-3}$ for $W+$jets background.
In Fig.~\ref{figwjj}, we present the $\missingET$ and
$M_{T}(l_1,\missingET)$ distributions
for signal and $W+$ jets background before VBF selections.
We can find that the $\missingET$ and $M_{T}(l_1,\missingET)$ cuts
can almost remove the $W+$jets background. Thus we neglect
it in the following analysis.
Based on the above discussion,
we summary our event selections in our final state analysis:
\begin{itemize}
 \item VBF selections: we require
at least four jets with $P_{T} > 40$ GeV in $\mid\eta\mid < 5$.
There must also be one pair of light jets $(j_1,j_2)$ satisfying;
(i) $\mid \Delta \eta (j_1,j_2) \mid > 4.2$ and $\eta_{j_1}\cdot\eta_{j_2} < 0$;
(ii) $P_{T}^{j_{1},j_{2}} > 50$ GeV; (iii) $M_{j_1j_2} > 500$ GeV.
 \item Lepton selection: only one lepton with $P_{T} > 20$ GeV and
  $\mid \eta \mid < 2.5$. We assume a $\tau$-tagging efficiency
  of 40$\%$ and include the mis-tagging of QCD jets in Delphes.
 \item $ \missingET > 150$ GeV.
 \item $M_{T}(\ell_1,\missingET) > 100$ GeV.
\end{itemize}

\begin{table}
\caption{The numbers of events for the signal
$p p \rightarrow \chi_1^\pm \chi_1^0 j j \rightarrow
 W^\pm h G^\prime G^\prime j j \rightarrow
 \ell^\pm \nu b \bar{b} G^\prime G^\prime j j \rightarrow
 \ell  + 4j + \missingET $ and backgrounds at the LHC with $\sqrt{s} = 14$ TeV and
 100 $fb^{-1}$ of integrated luminosity.}
\begin{tabular}{c||c|c|c|c||c}
 \hline \hline
\cline{1-6}
 cut &~~$t\bar{t}\rightarrow l\nu bbjj$~~  &  ~~$t \bar{t}\rightarrow l\nu l\nu bb$~~  &  ~~$tX \rightarrow l \nu b X$~~  &  ~~$WVjj \rightarrow l\nu jjjj$~~   & ~~signal~~ \\
 \hline
 VBF selctions  &  341917  &  49824  &  48512   &   32528  &  255   \\
  \hline
  Lepton selections  &  162975  &  22761  &  22389   &  15975   &  130   \\
 \hline
 $\missingET > 150$ GeV  &  11480    &   2603    &   1332  &   1836     &   53  \\
 \hline
 $M_{T}(\ell_1,\missingET) > 100$ GeV &  1731  &  1586  &  295   &  236  &  28   \\
 \hline
 \hline
\end{tabular}
\label{table1}
\end{table}

In Table~\ref{table1}, we present the numbers of events for signal and background
processes under the above cuts at the LHC with $\sqrt{s} = 14$ TeV and
100 $fb^{-1}$ of integrated luminosity. The VBF selections reduce the backgrounds
effectively, especially the top pair background. Although the light jet pair
that comes from $W$-boson decay in $t \bar{t}$ production could not pass
the VBF selections, bottom quarks or hadronic $\tau$ would be misidentified
as light jets so that the event could survive. So there are still many events
from the top pair backgrounds after VBF selections.
Table~\ref{table1} shows that the $\missingET$ cut is very
effective in reducing the backgrounds.
As we expected, a rather hard cut on $M_{T}(\ell_1,\missingET)$ could further
suppress the background and improve the significance.

\begin{table}
\caption{The statistical significance of the signal
$p p \rightarrow \chi_1^\pm \chi_1^0 j j \rightarrow
 W^\pm h G^\prime G^\prime j j \rightarrow
 \ell^\pm \nu b \bar{b} G^\prime G^\prime j j \rightarrow
 \ell  + 4j + \missingET $
at the LHC with $\sqrt{s} = 14$ TeV and different luminosities.
$S_{1}$ and $B_{1}$ stand for the signal and background events after VBF selection,
while $S_{2}$ and $B_{2}$ stand for the signal and background events after all the cuts.}
\begin{tabular}{c|c|c|c|c|c}
 \hline \hline
 ~~~$\sqrt{s} = 14$ TeV~~~  &  ~100 fb$^{-1}$~  &  ~500 fb$^{-1}$~  &  ~1000 fb$^{-1}$~  &  ~2000 fb$^{-1}$~  &  3000 fb$^{-1}$   \\
 \hline
 $S_{1}$/$\sqrt{ S_{1}+B_{1}}$  &  0.37  &  0.83  &  1.17  &  1.66  &  2.03   \\
 \hline
 $S_{2}$/$\sqrt{ S_{2}+B_{2}}$  &  0.46  &  1.02  &  1.45  &  2.04  &  2.5   \\
 \hline
 \hline
\end{tabular}
\label{table2}
\end{table}

In Table~\ref{table2} we display the signal significance for different luminosities
at the 14 TeV LHC.  As expected, the significance is improved by the cuts efficiently.
With a luminosity of 3000 $fb^{-1}$, a statistical significance of 2.5$\sigma$
can be achieved. We notice that the ratio of signal to backgrounds is very small,
which means that the systematic uncertainty should be well controlled in order to
detect the signal.

\subsection{Observability of  $\chi_1^+ \chi_1^- j j$ production  at the LHC}
Now we turn to the production of $\chi_1^+ \chi_1^- j j$
at the LHC. Since the chargino decays to a $W$-boson and a pseudo-goldstino,
the signal of this production is characterized by two opposite sign
leptons and a pair of forward/backward jets associated with large $\missingET$:
\begin{equation}
 p p \rightarrow \chi_1^+ \chi_1^- j j \rightarrow
 W^+ G^\prime W^- G^\prime j j \rightarrow
 \ell^+ \ell^- + 2j + \missingET  ~~(\ell = e,\mu, \tau).
\end{equation}
The dominant background comes from the top pair dileptonic processes.
As we discussed before, it could be reduced by VBF selections effectively.
The two opposite sign $W$-boson or $\tau$ production
associated with two jets can fake the signal,
where $W$ or $\tau$ decay leptonically.
In addition, another background comes from
$ZZ$ production associated with two jets, with one of the $Z$ bosons
decays to leptons and the other decays to neutrinos.

Since an important feature of the VBF processes is the absence of
color exchange between the forward/backward jets and this leads
to a suppression of hadron productions between these two jets,
we could enhance the signal to background ratio by vetoing addition
jets in the rapidity gap region between these jets. This cut will be
effective for suppressing the top  pair backgrounds. We also veto $b$-jets
to further suppress the top pair. The $\missingET$ in the backgrounds
comes from neutrinos from $W/Z$ boson or $\tau$ lepton decay.
But in the signal the pseudo-goldstino give rise to $\missingET$.
Therefore a large $\missingET$ cut and $M_{T}(\ell_1,\missingET)$
will reduce all the backgrounds and improve the signal significance.
In summary, we employ the following cuts
\begin{itemize}
 \item VBF selections: we require a pair of light jets $(j_1,j_2)$ satisfying
(i) $\mid \Delta \eta (j_1,j_2) \mid > 4.2$
and $\eta_{j_1}\cdot\eta_{j_2} < 0$;
(ii)  $P_{T}^{j_{1},j_{2}} > 50$ GeV;
(iii) $M_{j_1j_2} > 500$ GeV.
\item Central jet veto: no jets with $P_{T}>$ 20 GeV between
$\eta_{j_1}$ and $\eta_{j_2}$.
 \item Lepton selection:  two opposite sign leptons with $P_{T} > 20$ GeV and
  $\mid \eta \mid < 2.5$.
 \item Veto $b$-jet: we reject events with any $b$-tagging jets. Note that we apply the
$b$-jet tagging and $c$-jet mis-tagging efficiency as in \cite{b-tagging} which
also includes a misidentified rate for the light jets.
 \item $ \missingET > 150$ GeV.
 \item $M_{T}(\ell_1,\missingET) > 100$ GeV.
  \end{itemize}

\begin{table}
\caption{The numbers of events for the signal $p p \rightarrow \chi_1^+ \chi_1^- j j \rightarrow
 W^+ G^\prime W^- G^\prime j j \rightarrow
 \ell^+ \ell^- + 2j + \missingET$ and backgrounds at the LHC with $\sqrt{s} = 14$ TeV and
 1000 $fb^{-1}$ of integrated luminosity.}
\begin{tabular}{c||c|c|c|c||c}
 \hline \hline
\cline{1-6}
 cuts & ~~$t \bar{t}\rightarrow l\nu l\nu bb$~~  & ~~~ $\tau \tau jj$ ~~~
&  $WWjj \rightarrow \ell\nu \ell\nu jj$  &  $ZZjj \rightarrow \ell\ell\nu\nu jj$ & ~~signal~~ \\
 \hline
 VBF selctions  &  912261  &  1852740  &  57330   &   2867  &  601   \\
 \hline
 Central Jet veto  &   54978  &  310074  &  11662  &     869    &  116     \\
  \hline
  Lepton selections  &  16606  &  22148  &  4606   &  441   &  55   \\
  \hline
  Veto b-jet  &    14287      &   22148  &  4606   &    441     &    55     \\
 \hline
 $\missingET > 150$ GeV  & 1272 &   963  &  588  &   135   &   25  \\
 \hline
 $M_{T}(\ell_1,\missingET) > 100$ GeV &  898  &  481  &  441 &  131  &   21   \\
 \hline
 \hline
\end{tabular}
\label{table3}
\end{table}

A summary of events with the luminosity of $1000 fb^{-1}$ at each
selection stage is displayed in Table~\ref{table3}.
The VBF cuts and central jet veto are very effective in reducing the backgrounds,
especially the top pair background. We also find that the large
$\missingET$ cut can suppress the important backgrounds to 1/11-1/23.
The $M_{T}(\ell_1,\missingET)$ cut could further reduce the backgrounds and
improve the signal significance.
The result shows that the significance ($S/\sqrt{S+B}$) can reach
about $0.48\sigma$ and $0.83\sigma$
for an integrated luminosity of 1000$fb^{-1}$ and 3000 $fb^{-1}$ at the 14 TeV LHC.
So we conclude that the signal  $p p \rightarrow \chi_1^+ \chi_1^- j j \rightarrow
 W^+ G^\prime W^- G^\prime j j \rightarrow
 \ell^+ \ell^- + 2j + \missingET$ is not accessible at the  14 TeV LHC.

\subsection{Pseudo-goldstino mass effects}
In our above study we simply assumed the pseudo-goldstino is massless.
Actually in concrete models with multi-hidden sectors, the pseudo-goldstino 
acquires a universal mass at tree level~\cite{Cheung:2010mc,Cheung:2011jq}, 
which is twice the gravitino mass, and also gets 
model-dependent contributions at loop level.  
Some authors have argued that the loop contributions should be at least 
the GeV scale \cite{Argurio:2011hs} (a concrete calculation is still missing 
even in the simplest model).
In the following we show the mass effects of the pseudo-goldstino 
in our simulations.

Comparing to a massless pseudo-goldstino, the phase space and the amount of missing transverse energy
for a massive  pseudo-goldstino will be reduced.
For the VBF processes with two energetic jets, the missing transverse energy should not be as
sensitive to the  pseudo-goldstino mass as the Drell-Yan processes.
To show this explicitly, we take the signal process $\chi^{\pm}\chi^{0}jj$ as an example.
We simulate the signal $p p \rightarrow \chi_1^\pm \chi_1^0 j j \rightarrow
 W^\pm h G^\prime G^\prime j j \rightarrow
 \ell^\pm \nu b \bar{b} G^\prime G^\prime j j$ with the pseudo-goldstino mass
of 40 GeV and 80 GeV, respectively.

\begin{figure}[htbp]
\includegraphics[scale=0.35]{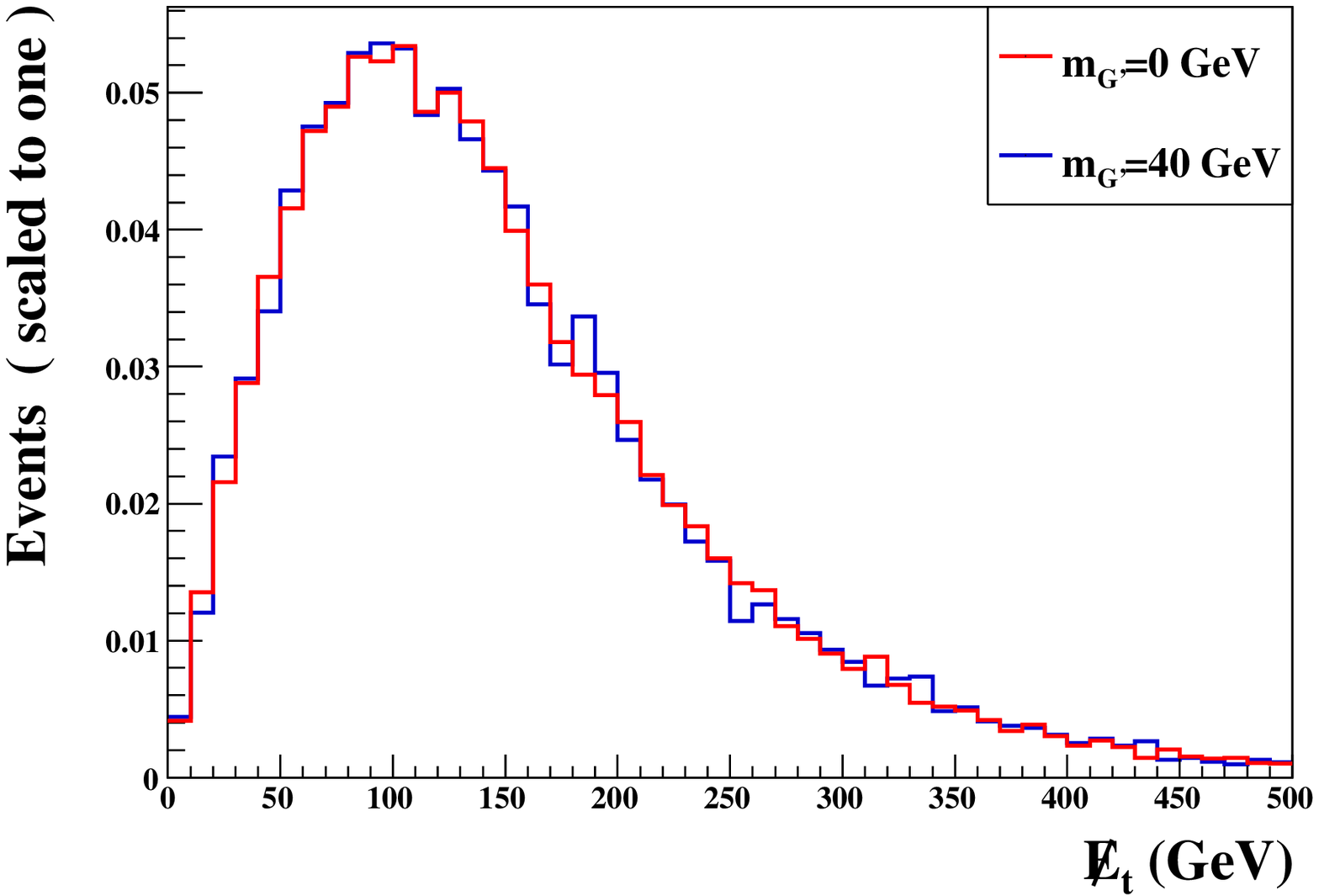}
\includegraphics[scale=0.35]{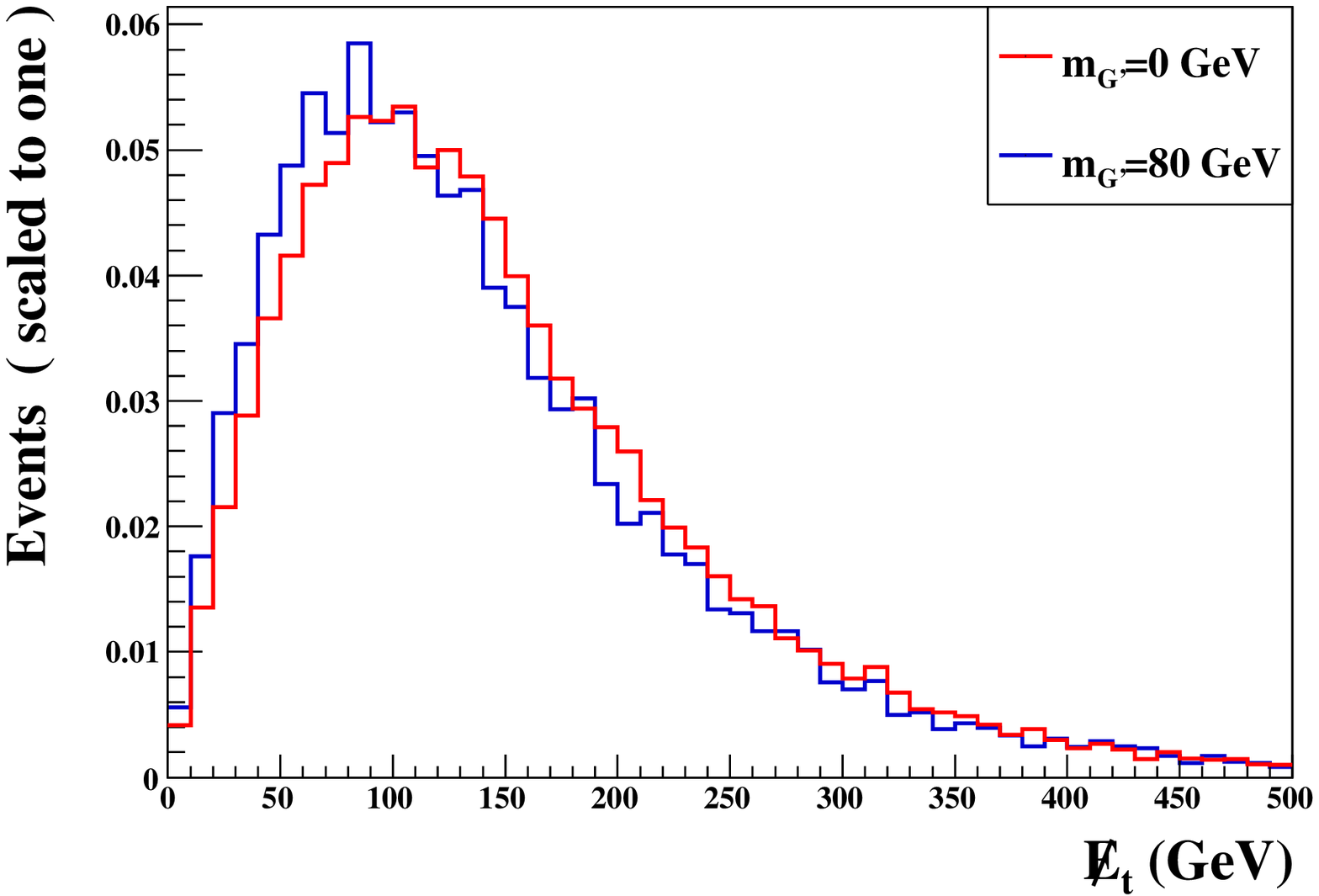}
\vspace*{-0.3cm}
\caption{The $\missingET$ distributions after VBF cut for the signal
$p p \rightarrow \chi_1^\pm \chi_1^0 j j \rightarrow
 W^\pm h G^\prime G^\prime j j \rightarrow
 \ell^\pm \nu b \bar{b} G^\prime G^\prime j j$ at the 14 TeV LHC
for different psedo-goldstino masses.}
\label{figmass}
\end{figure}

In Fig.~\ref{figmass} we present the $\missingET$
distribution after VBF cut for a massive pseudo-goldstino compared with the massless
case. From the left panel we can see that the $\missingET$ distributions with $m_{G^{\prime}}=0,40$ GeV
almost overlap with each other. The $\missingET$ distribution with $m_{G^{\prime}}=$80 GeV
is presented in the right panel of Fig.~\ref{figmass}, which shows that
the missing transverse energy is just a little softer than in the massless case.
Note that 80 GeV is the largest value of pseudo-goldstino mass to open the neutralino decay channel.
Finally, we also present the signal events with 100 $fb^{-1}$ of integrated luminosity for
different pseudo-goldstino masses in Table~\ref{tablemass}. We see that as the pseudo-goldstino
mass increases, the efficiency of the $\missingET$ cut slightly decreases. The results also show
that the signal event number is not sensitive to the pseudo-goldstino mass.
In the last low of Table~\ref{tablemass}, we display
the signal significances for different pseudo-goldstino masses with 3000 $fb^{-1}$
of integrated luminosity at the 14 TeV LHC. The significance
can reach to 2.2$\sigma$ and 1.9$\sigma$ with $m_{G^{\prime}}=$40 GeV and 80 GeV,respectively.

\begin{table}
\caption{The numbers of events for the signal
$p p \rightarrow \chi_1^\pm \chi_1^0 j j \rightarrow
 W^\pm h G^\prime G^\prime j j \rightarrow
 \ell^\pm \nu b \bar{b} G^\prime G^\prime j j$ with different pseudo-goldstino
 masses for 100 $fb^{-1}$ of integrated luminosity. The signal significance
 is shown for 3000 $fb^{-1}$ of integrated luminosity at the LHC with $\sqrt{s} = 14$ TeV.}
\begin{tabular}{c||c|c|c|c||c}
 \hline \hline
\cline{1-2}
 cut & VBF selctions  &  Lepton selections  & $\missingET > 150$ GeV  &  $M_{T}(\ell_1,\missingET) > 100$ GeV  & $S/\sqrt{S+B}$  \\
 \hline
 $m_{G^{\prime}}=0$ GeV &  255  &  130  &  53   &   28  & 2.5  \\
 \hline
 $m_{G^{\prime}}=40$ GeV &  252  &  128  &  51   &   25  & 2.2  \\
 \hline
 $m_{G^{\prime}}=80$ GeV &  243  &  122  &  45   &   22  &  1.9 \\
 \hline
 \hline
\end{tabular}
\label{tablemass}
\end{table}

Next, we look at the electroweakino search results at the LHC.
The most relevant search is the production of chargino
and neutralino in final states with $l+b\bar{b}+\missingET$ by
ATLAS \cite{atlas} and CMS\cite{cms}.
They have interpreted their results in the context of a simplified model,
where they assumed the lightest neutralino ($\chi_{1}^{0}$) is bino-like while
the second lightest neutralino ($\chi_{2}^{0}$) and the lightest chargino
($\chi_1^{\pm}$) are wino-like (their masses are approximately degenerate
$m_{\chi_1^{\pm}}=m_{\chi_{2}^{0}}$). They searched the production $pp \rightarrow \chi_1^{\pm}
\chi_2^{0} \rightarrow (W^{\pm}\chi_1^{0})(h \chi_1^{0})$ (with 100$\%$ branching
ratios) by employing the $h \rightarrow b\bar{b}$ channel. The limit was found to
be $m_{\chi_2^{0}}=m_{\chi_1^{\pm}}\gtrsim$ 200 GeV \cite{cms}
and 300 GeV \cite{atlas} for $m_{\chi_1^{0}} \lesssim$ 30 GeV.
It means that in our scenario the benchmark point $m_{\chi_1^{\pm}}=m_{\chi_1^{0}}
=200$ GeV may have been excluded for $m_{G^{\prime}} \lesssim 30$ GeV and
survive for $m_{G^{\prime}} = 40$ GeV and 80 GeV.

Before concluding, we make some comments.
(1) Our simulations above are just to demonstrate that our scenario
are possiblly accessible at the high-luminosity LHC. More dedicated selections of
the signal from the backgrounds may improve the significance, which should be
considered in the future experimental search. (2) In our study we focused on MSSM,
while in other low energy SUSY models the pseudo-goldstino may also have similar
new decay channels which deserve searches.

\section{CONCLUSION}
The multi-sector SUSY breaking scenario predicts pseudo-goldstino,
which can couple to the visible sector more strongly than the
ordinary gravitino. Then the lightest electroweakinos can decay to a
pseudo-goldstino plus a $Z$-boson, Higgs boson or $W$-boson.
In our previous work \cite{Hikasa:2014yra} we investigated the
Drell-Yan productions of the lightest electroweakinos followed by the
decays to pseudo-goldstino at the LHC. In this work we extended the
study to the VBF productions of  electroweakinos.
From the Monte Carlo simulations
we found that the largest rate channel $pp\to \chi_1^{\pm} \chi_1^0 jj$
can have a statistical significance above  $2\sigma$ at the 14 TeV LHC
with an luminosity of 3000 fb$^{-1}$, while the second largest rate
channel $p p \rightarrow \chi_1^+ \chi_1^- j j$ is not accessible.

Finally we point out that in our study we considered
the decays of electroweakinos to a pseudo-goldstino, i.e.,
$\chi_1^+\to W +G'$ and  $\chi_1^0\to h + G'$. Our results are approximately
applicable to other scenarios which predict a light singlet invisible
particle ($X$) as long as the decays $\chi_1^+\to W + X$ and  $\chi_1^0\to h + X$
happen inside the detector. For example, in the next-to-minimal supersymmetric model,
the singlino-like lightest neutralino can be as light as a few GeV \cite{cao} and
thus similar decays can happen for the chargino/neutralino.

\section*{Acknowledgments}
Lin Wang acknowledges Prof.\ Johann K\"{u}hn and Prof.\ Matthias Steinhauser
for their warm hospitality. We thank Lei Wu for discussions.
This work is supported by
by DFG through SFB/TR 9
``Computational Particle Physics'' and
by the National Natural
Science Foundation of China under grant Nos. 11275245, 10821504 and
11135003.

\end{document}